\documentclass[superscriptaddress,aps,showpacs,nofootinbib,twocolumn]{revtex4}
\usepackage{graphicx}
\usepackage{amsmath,amssymb}

\def\CQG{{\it Class. Quantum Gravity} }
\def\EURO{{\it The European Phys. J.} }

\def\NAT{{\it Nature} }

\def\PR{{\it Phys. Rev.} }
\def\PRL{{\it Phys. Rev. Lett.} }

\def\vev#1{\langle {#1}\rangle}

\def\frac#1#2{{\textstyle{{#1}\over {#2}}}}

\def\lsim{\mathrel{\rlap{\lower4pt\hbox{\hskip1pt$\sim$}}
    \raise1pt\hbox{$<$}}}
\def\gsim{\mathrel{\rlap{\lower4pt\hbox{\hskip1pt$\sim$}}
    \raise1pt\hbox{$>$}}}
\def\sqr#1#2{{\vcenter{\vbox{\hrule height.#2pt
         \hbox{\vrule width.#2pt height#1pt \kern#1pt
         \vrule width.#2pt}
         \hrule height.#2pt}}}}

\def\beq{\begin{equation}}
\def\eeq{\end{equation}}
\def\beqa{\begin{eqnarray}} 
\def\eeqa{\end{eqnarray}}

\def\laq{\raise 0.4 ex \hbox{$<$}\kern -0.8 em\lower 0.62 ex\hbox{$\sim$}}
\def\gaq{\raise 0.4 ex \hbox{$>$}\kern -0.7 em\lower 0.62 ex\hbox{$\sim$}}

\begin{document}

\title{Quantum and Classical divide: the gravitational case}

\author{O. Bertolami}
\altaffiliation{Email address: orfeu@cosmos.ist.utl.pt}

\author{J. G. Rosa}
\altaffiliation{Email address: joaopedrotgr@sapo.pt}
 
\affiliation{ Departamento de F\'\i sica, Instituto Superior T\'ecnico \\
Avenida Rovisco Pais 1, 1049-001 Lisboa, Portugal}

\vskip 0.5cm

\date{\today}

\begin{abstract}

We study the transition between quantum and classical behaviour of particles in a gravitational quantum well. We analyze how an increase in the particles mass turns the energy spectrum into a continuous one, from an experimental point of view. We also discuss the way these effects could be tested by conducting experiments with atoms and fullerene-type molecules.

\end{abstract}

\pacs{03.65.-w, 36.40.-c \hspace{2cm}Preprint DF/IST-5.2005} 

\maketitle
 
\section{Introduction}

The understanding of the conditions of the transition between the quantum and the classical descriptions is a 
central question in Quantum Mechanics. The issue has repeatedly been discussed since Schr\"odinger's cat first 
appeared in the literature \cite{Schrodinger}. More recently, this somewhat conceptual and philosophical discussion 
has become a quite concrete experimental problem given the substantive progress one has witnessed on the question of 
loss of quantum coherence, usually referred to as decoherence \cite{Zureck} and the interference experiments with 
macromolecules \cite{Arndt, Hornberger, Hackermuller}. These experiments have, in particular, examined the conditions 
under which the quantum coherence is lost for highly complex systems. It is believed that determining the drawing line 
between quantum and classical behaviour may bring new insights into the nature of macroscopic objects which exhibit 
quantum properties \cite{Leggett}, and may allow spelling out the conditions under which, for instance, quantum computers 
may be physically built. 

In this work, we study the transition between the quantum and classical regimes
in a Gravitational Quantum Well (GQW). In such a system, particles exhibit a
discrete energy spectrum and present a non-vanishing probability of tunneling
into classically forbidden regions. We analyze how an increase in the particles
mass may destroy these quantum properties, so that the system will, at least
from an experimental point of view, behave as its classical analogue. Thus, we
propose a generalization of the experiment performed by Nesvizhevsky \emph{et
al.} \cite{Nesvizhevsky_1,Nesvizhevsky_2} with neutrons to more massive
particles, such as atoms and fullerene-type molecules.

This new criteria is based on the assumption that the dependence of separation between quantum 
states is a reliable criteria for establishing the quantum to classical divide, at least within 
the achievable experimental resolution. In this work we consider atoms and fullerene-type molecules. We shall restrict ourselves 
to the GQW in Quantum Mechanics. An extension of this system to, for instance, a noncommutative phase space has been 
studied in Refs. \cite{Bertolami_1,Bertolami_2}. The Equivalence Principle in the context of the GQW has been discussed in Ref. 
\cite{Bertolami_3}.

\section{The Gravitational Quantum Well}

Consider a particle of mass M moving on the $xy$ plane in a uniform gravitational field $\mathbf{g}=-g\mathbf{e_x}$. When a horizontal mirror is placed at $x=0$, a \emph{gravitational quantum well} is set up. In the direction transverse to the gravitational field, $y$, the particle is free, exhibiting a continuous energy spectrum. In the direction of the gravitational field, $x$, the particle exhibits a discrete energy spectrum. In the $n$-th quantum level, the particle's wave function is given by the Airy function $\phi(\xi_n)$, where $\xi_n=(x-x_n)/x_0$ and $x_0\equiv(\hbar^2/2M^2g)^{1/3}$ \cite{Landau}. The energy eigenvalues are given by:
\beq \label{GQW_2}
E_n=-\bigg({Mg^2\hbar^2\over2}\bigg)^{1/3}\alpha_n~,
\eeq
where $\alpha_n$ represents the $n$-th zero of the Airy function. The value of $x_n=E_n/Mg=-x_0\alpha_n$ corresponds to the maximum height which is classically allowed for a particle with energy $E_n$.

The probability of finding the particle is non-vanishing for all values of $x>0$. However, as soon as $x$ exceeds the value of $x_n$ for each quantum state $n$, this probability decays exponentially. Nevertheless, the particle has a finite probability of penetrating a classically forbidden region through quantum tunneling.

This idea has been used to study the spectrum of neutrons in the quantum well of the Earth's gravitational field and a horizontal mirror \cite{Nesvizhevsky_1, Nesvizhevsky_2}. An ultra cold neutron beam is considered, with a mean velocity $v=6.5 \ \mathrm{ms^{-1}}$, traveling through a narrow slit formed by the mirror and a scatterer/absorber placed above it. Then, the neutron flux through the apparatus is measured as a function of the slit height, $x$. For $x>x_n$, neutrons in the $n$-th quantum state have a small probability of crossing the gravitational barrier and tunnel into the scatterer/absorber. This probability is approximately given by $\exp(-4/3\xi_n^{3/2})$ and quickly vanishes as the slit height increases. Hence, for $x>x_n$, neutrons pass through the slit with little loss. For $x<x_n$, however, neutrons have a $O(1)$ probability of being absorbed by the scatterer and so the slit is not transparent to neutrons. The flux of neutrons through the slit is, thus, given by:
\beqa \label{GQW_4}
F(x)=F_0\sum_n\beta_ne^{-{L\omega_n\over v}\left\{\begin{array}{ll}
e^{\big(-{4\over3}\xi_n^{3/2}\big)}\ ,\xi_n>0\\
1\qquad\qquad ,\xi_n<0
\end{array}\right\}}~,
\eeqa
where $F_0$ is a normalization factor, depending on the incident flux, $\beta_n$ is the relative population of the $n$-th quantum level, $L$ is the length of the slit and $\omega_n\equiv(E_{n+1}-E_n)/\hbar$ \cite{Nesvizhevsky_2}. From Eq. (\ref{GQW_4}), one can conclude that, for $x\gg x_n$, the flux of particles in 
the $n$-th state approaches its maximum value $F_0$, while for $x\ll x_n$, this flux tends to zero. Hence, the classical turning points $x_n$ separate two regions where the neutron flux exhibits two distinct types of behaviour for each state $n$. By considering a model of this type and adjusting the experimentally measured flux to the predicted flux given by Eq. (\ref{GQW_4}), the values of the two lowest classical turning points were obtained \cite{Nesvizhevsky_2}, 
which are in good agreement with the quantum mechanical predictions, $x_1=13.7\ \mathrm{\mu m}$ and $x_2=24.0\ \mathrm{\mu m}$.


\section{The Classical Limit}

The GQW is a convenient system for testing whether a system behaves as a quantum or a classical system, as its energy spectrum depends on the mass of the particles involved. In the following, we shall analyze the consequences of increasing this mass and consider its experimental implications.

We point out that, for all particles, the energy spectrum approaches a classical continuous spectrum for high energy values. This can be seen by analyzing the asymptotic form of the zeros of the Airy function \cite{Fabijonas}:
\beq \label{Airy_zeros_1}
\alpha_n=-T\bigg({3\over8}\pi(4n-1)\bigg)~,
\eeq
where $T(t)\approx t^{2/3}$ for $t\gg1$. Hence, for $n\gg1$,
\beq \label{Airy_zeros_2}
\Delta\alpha_n\equiv|\alpha_{n+1}-\alpha_n|\approx{2\over3}\bigg({3\over2}\pi\bigg)^{2/3}n^{-1/3}
\eeq
Thus, both $\Delta E_n\equiv E_{n+1}-E_n$ and, consequently, $\Delta_n\equiv x_{n+1}-x_n$, tend to zero as $n\rightarrow\infty$. Also, $\Delta\alpha_n$ is strictly decreasing with $n$, and so the largest separation between consecutive heights $x_n$ occurs for the first two quantum states. 

Let us now analyze the effects of increasing the particle's mass. Observing Eq. (\ref{GQW_2}), one can conclude that all the energy eigenvalues are proportional to $M^{1/3}$ and the same occurs for the separation between consecutive energy eigenvalues. This is somewhat unexpected, as one assumes that this separation decreases with increasing mass. However, one should notice that the experimentally relevant quantities are the heights $x_n$. One can easily conclude that these heights, as well as the separation $\Delta_n$, are proportional to $M^{-2/3}$. Thus, for more massive particles, it will become harder to distinguish consecutive levels. If two of these cannot be separated experimentally, one has no means of distinguishing the quantum states and the spectrum appears classical.

As the largest separation between values of $x_n$ occurs for the two lowest quantum states, then if the particle's mass is large enough so that the experimental error is larger than $\Delta_1$, there will be no way of separating any two consecutive classical turning points. In this case, the flux of particles through the slit will exhibit a classical behaviour, increasing as $x^{3/2}$ \cite{Nesvizhevsky_2}. Hence, classical or quantum behaviour depend on the mass of the particles under study and on the experimental resolution. We quantify this in the following way: if there is a minimum uncertainty, $\epsilon$, for measuring the slit height, $x$, then one can distinguish at least two consecutive quantum states if the mass of the particles does not exceed:
\beq \label{limit_1}
M_{max}=\sqrt{{\hbar^2\over2g}}\bigg({\Delta\alpha_1\over\epsilon}\bigg)^{3/2}~.
\eeq

For instance, with the error of $\epsilon=2.5\ \mathrm{\mu m}$ for $n=1$ in \cite{Nesvizhevsky_2}, one could distinguish the first two quantum states for particles with a mass $M\lesssim 8m_N$, where $m_N=1.67\times10^{-27}\ \mathrm{kg}$ denotes the mean mass of a nucleon. 

Equivalently, in order to distinguish at least two quantum states of a particle with mass $M=Am_N$, one must have a maximum experimental error given by:
\beq \label{limit_2}
\epsilon_{max}=\bigg({\hbar^2\over2m_N^2g}\bigg)^{1/3}\Delta\alpha_1A^{-2/3}\simeq 10.3 A^{-2/3}\ \mathrm{\mu m}~.
\eeq
Furthermore, it has to be taken into account the fact that the Uncertainty Principle places limits on the experimental resolution for particles with finite lifetimes. Measuring $x_n$, an uncertainty $\Delta x_n$ is equivalent to an uncertainty $\Delta E_n=Mg\Delta x_n$\footnote{This follows from the 
variational computation: $\delta E_n = \delta (<n|\hat{H}|n>)=<n|\delta \hat{H}|n>=<n|Mg\delta \hat{x}|n>=Mg<n|\delta \hat{x}|n>=Mg\delta x_n$.}. Hence, in order to have enough spatial resolution to separate the first two quantum states of a particle with mass $M=Am_N$, its mean lifetime must be greater than:
\beq \label{limit_3}
\Delta\tau_{min}=\bigg({2\hbar\over m_Ng^2}\bigg)^{1/3}{A^{-1/3}\over\Delta\alpha_1}\simeq 6.3\times10^{-4}A^{-1/3}\ \mathrm{s}~.
\eeq
For the Nesvizhevsky \emph{et al.} set up, the minimum lifetime is $0.63$ ms, which is 6 orders of magnitude smaller than the neutron's mean lifetime of 885.7 s \cite{PDG}. The Uncertainty Principle allows a precision up to $10^{-5}\ \mathrm{\mu m}$ and, thus, the experiment can be further improved.

Another consequence of increasing the mass of the particles in the GQW is the decrease in the probability of tunneling through the gravitational barrier. As already referred to, in state $n$ this probability is approximately given by $\exp(-4/3\sqrt{2g/\hbar^2}(x-x_n)M)$.  This rapid decrease is expected, as in the classical limit particles with energy $E_n$ cannot be found at a height larger than $x_n$. Of course the frequency $\omega_n$, which can be viewed as the frequency of the collisions between the particles and the gravitational barrier \cite{Nesvizhevsky_2}, increases as $M^{1/3}$. but this does not compensate the decrease of the tunneling probability. Thus, more massive particles exhibit a sharper transition between zero and maximum flux, tending to a discontinuous transition in the classical limit $M\rightarrow+\infty$. This makes it harder to observe distinct quantum states.


\section{Finite size effects: Atoms and Fullerene Molecules}

So far, we have only dealt with point particles, neglecting the effects of their size. For neutrons, the average radius of $\simeq1\ \mathrm{fm}$ is about 10 orders of magnitude smaller than the correspondent value of $x_1$ and so the former may be neglected. However, as the particles mass increases, their size has to be taken into account.

Let us then analyze how size affects the GQW energy spectrum. 
Before that we point out that although the GQW problem has only been solved for point-like particles,
some of the consequences of the particles finite size may still be inferred
from a qualitative analysis of the problem. For that, let us suppose that the effects of gravity on all the elementary particles which constitute a massive system are much smaller than the effects of the forces which bound them. With this hypothesis, one can focus on the movement of the massive particle's center of mass (CM), which follows the same rules as in the point particle approximation. The most important difference between these two descriptions is that the wave function of a composed particle exhibits an additional dispersion around the position of its CM. This dispersion is quantified by the particle's mean radius, $R$. Thus, in a GQW experiment, when the slit height becomes of order $x_n+R$, the scatterer/absorber gets close enough to the wave function of a particle in the $n$-th state so that the latter can be absorbed at $x_n + R$, the CM position. Hence, the transition between the regions of maximum and minimum flux occurs in the neighborhood of $x_n+R$ and not of $x_n$, as in the point particle approximation. When $R\gtrsim x_n$, this must be taken into account.

Another consequence of the finite radius is that massive particles will tend to leave the lowest quantum states less populated. The horizontal mirror implies the wave function of each of the elementary particles that constitute the composed one vanishes at the origin. This means that the whole composed particle must be above the horizontal mirror. Hence, the particle's CM must be at a height larger than its average radius $R$. If $R>x_n$, particles in the $n$-th level are only allowed to be in the classically forbidden region, where they have a very low probability of being found. Thus, the great majority of particles will be found in quantum states $n+1$ and higher. The more massive the particle the more likely it will lie at higher quantum states. This solves an apparent paradox of our earlier discussion, namely that the growth in the particle mass implies an increase in the separation between its energy levels. However, as its radius also increases, the particle will more probably be found in the higher quantum levels, where $\Delta E_n\rightarrow0$.

One must also bear in mind that, as the whole particle must lay above the horizontal mirror, there is a minimum value of the slit height, corresponding to the average diameter of the particle, that can be experimentally tested, as for smaller values the particles cannot fit in the narrow space between the mirror and the scatterer/absorber. For $2R$ of order $x_n+R$, one is no longer able to observe the transition between zero and maximum flux. This occurs when $R\sim x_n$, so that one stops being able to measure the $n$-th quantum state at the same time the particle stops being able to be found on it. Hence, one is not able to test experimentally whether particles of radius $R\sim x_n$ are not really in the $n$-th quantum state, unless one may extrapolate the population of this level from the population of the higher levels.

We consider now the properties of the particles that may be used to test the quantum to classical transition in the GQW. There are some features these particles must possess:
(i) They must be electrically neutral so that neither electromagnetic effects overlap the gravitational ones nor the decoherence of the particle beam is induced. A highly symmetric spatial distribution of electrons prevents polarization effects. Having vanishing total spin avoids the coupling to external magnetic fields\footnote{Although the interaction of particles with the mirror can 
pose some problems.};
(ii) They must have long lifetimes, so to satisfy Eq. (\ref{limit_3}). Thus, radioactive particles are not appropriate;
(iii) The horizontal velocity of the beam must be small in order to maximize the time the particles remain on the GQW states.

With these requirements, atoms are, after neutrons, the natural choice in the mass hierarchy. According to (i) and (ii), they must not be in ionized states and must be stable isotopes with valence electrons in $s$-type orbitals, which have zero orbital angular momentum. The main contribution to an atom's mass comes from its nucleons, so $M=Am_N$ and from Eqs. (\ref{limit_2}) and (\ref{limit_3}) one computes $\epsilon_{max}$ and $\Delta\tau_{min}$. The former has been plotted in Figure 1 for $A<88$ (lanthanides and actinides are excluded). From these, one concludes that the maximum allowed error lies in the interval $0.1-10\ \mathrm{\mu m}$. The minimum lifetime can be shown to lie in the interval $0.1-0.63$ ms.
\begin{figure}[htbp]
	\centering
		\includegraphics[scale=0.75]{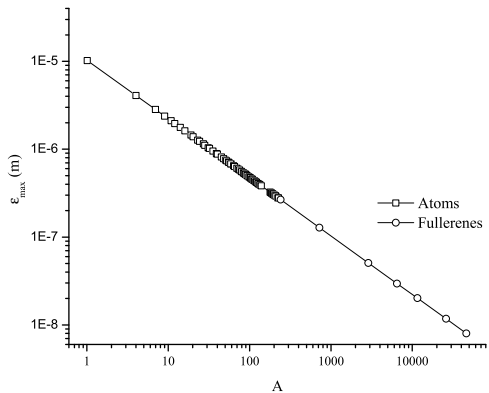}
	\caption{Maximum error for atoms and fullerene molecules.}
	\label{fig:Erro}
\end{figure}

The atomic radius can be estimated using the average value of the radial coordinate for a hydrogen type atom \cite{Gasiorowicz}, replacing the atomic number, $Z$, by an effective one, $Z_{eff}$, which takes into account the shielding effects of the inner electrons according to the Slater rules \cite{Slater}:
\beq \label{r_atoms}
R\simeq\vev{r}={a_0\over2Z_{eff}}[3n^2-l(l+1)]~,
\eeq
where $a_0\simeq5.29\times10^{-11}$ m is the Bohr radius and $(n,l)$ denote the atom's valence orbital quantum numbers. In Figure 2 the values of the ratio $R/\Delta_1$ for atoms with $A<88$ are plotted, and one can see that they lie in the interval $10^{-5}-10^{-3}$.
\begin{figure}[htbp]
	\centering
		\includegraphics[scale=0.75]{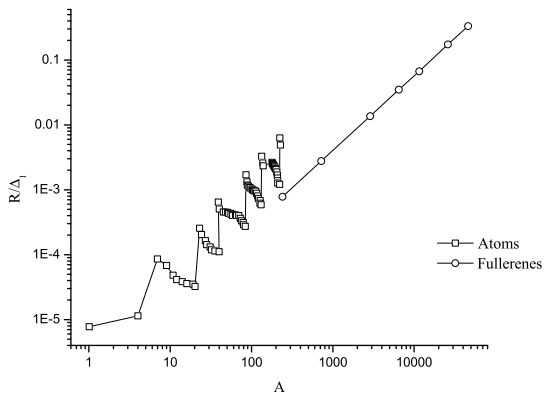}
	\label{fig:Fraccao}
	\caption{Ratio $R/\Delta_1$ for atoms and fullerene molecules.}
\end{figure}

Advancing in the mass hierarchy, we find molecules. There is a broad variety of molecules, but few satisfy our criteria. Among these, fullerenes or \emph{buckyballs}, i.e. large groups of $_6^{12}\mathrm{C}$, arise as the best candidates for our GQW criteria. These molecules are highly symmetric ($\mathrm{C}_{60}$ has spherical symmetry) and are globally neutral, although with a slight polarization \cite{Gueorguiev}. They are much bigger than atoms and the higher number of correlated particles enhances the probability of decoherence of the beam through interactions with the external environment. Furthermore, they have a large number of internal degrees of freedom (rotational and vibrational) and can radiate, thus interacting with the external environment. Nevertheless, they are more appropriate than most massive molecules for the GQW study.

For pure $_6^{12}\mathrm{C}$ fullerenes, $A=12N_C$, where $N_C$ is the number of Carbon atoms. The values of $\epsilon_{max}$ and $\Delta\tau_{min}$ for fullerenes up to 3840 Carbon atoms (see numerical simulations in Ref. \cite{Gueorguiev}) were computed using Eqs. (\ref{limit_2}) and (\ref{limit_3}). The values of $\epsilon_{max}$ are plotted in Figure 1. One finds that in order to distinguish at least the first two quantum states of these molecules in a GQW the maximum error must lie in the interval $0.01-0.1\ \mathrm{\mu m}$, corresponding to a minimum lifetime of 0.02 to 0.10 ms. 

If one assumes a fullerene to be modeled by a sphere of radius $R$, than the area of the sphere must proportional to the number of Carbon atoms \cite{Park}, $R=k\sqrt{N_C}$, where the constant $k\simeq4.38\times10^{-11}$ m is determined using values of \cite{Gueorguiev}. The ratio $R/\Delta_1$ for fullerenes is plotted in Figure 2 and lies in the interval $10^{-3}-10^{-1}$. Thus, one sees that the finite size effects cannot be neglected for the largest fullerenes. One can estimate the value of $N_C$ for which $R$ becomes of the order of $x_1$, concluding that molecules with more than $\sim 12\:470$ carbon atoms will have a very small probability of being in the lowest quantum state of the GQW. Hence, the considered fullerenes ($N_C<3840$) can be found in the first quantum level.


\section{Conclusions}

In this paper we have studied the consequences of the increase in the particles mass and size on the energy spectrum of the GQW. We have found that it leads to a decrease in the separation of consecutive classical turning points. Hence, the more massive the particles, the bigger the precision required to distinguish the lowest two quantum states. The precision already achieved allows studying particles up to $A\sim8$. For heavier atoms, it is necessary to increase the precision of position measurements at least by a factor of 10. For fullerenes up to $\mathrm{C}_{60}$, this increase in precision may be sufficient, but for larger molecules such as $C_{3840}$ an improvement of at least a factor of 1000 is required 

Of course, performing this kind of experiment with sizable particles may bring many experimental challenges. Besides the discussed precision, one requires strong isolation from external agents such as electromagnetic fields, although this 
may not be sufficient due to the presence of the mirror, and precautions to avoid collisions so not to decohere of the particle beam. Ultra cold beams of particles are also needed in order to maximize the time the particles are captured in the GQW.

Despite these difficulties, performing this kind of experiment might be of great importance for testing the limits of applicability of the Quantum Mechanics. In the suggested set up, the transition to a classical regime is made independently of the phenomenon of decoherence, depending only on the mass of the particles in the GQW and on the experimental resolution. Observing quantum states of massive particles in a GQW may turn out to be an important complement to quantum interference experiments with complex molecules \cite{Arndt, Hornberger, Hackermuller}. 

\vskip 0.2cm

\centerline{\bf {Acknowledgments}}

\vskip 0.2cm

\noindent 
One of us (O.B.) would like to thank Prof. A. Leggett for an illuminating 
discussion on the limits of Quantum Mechanics.  

\vfill


\end{document}